\documentclass[aps,prl,twocolumn,superscriptaddress,amsmath,amsfonts,emails,
               floatfix]{revtex4}

\usepackage{epsfig,psfrag}
\usepackage{CJK}

\usepackage{graphicx}
\usepackage{dcolumn}
\usepackage{bm}
\usepackage{amssymb}

\begin{document}
\title{Self-trapped interlayer excitons in van der Waals heterostructures}
\author{Jia-Pei Deng}
\affiliation{Tianjin Key Laboratory of Low Dimensional Materials Physics and Preparing Technology, Department of Physics, School of Science, Tianjin University, Tianjin 300354, China}
\author{Hong-Juan Li}
\affiliation{College of physics and Intelligent Manufacturing Engineering, Chifeng University, Inner
Mongolia, Chifeng 024000, China}
\author{Xu-Fei Ma}
\affiliation{Tianjin Key Laboratory of Low Dimensional Materials Physics and Preparing Technology, Department of Physics, School of Science, Tianjin University, Tianjin 300354, China}
\author{Xiao-Yi Liu}
\affiliation{Tianjin Key Laboratory of Low Dimensional Materials Physics and Preparing Technology, Department of Physics, School of Science, Tianjin University, Tianjin 300354, China}
\author{Yu Cui}
\affiliation{Tianjin Key Laboratory of Low Dimensional Materials Physics and Preparing Technology, Department of Physics, School of Science, Tianjin University, Tianjin 300354, China}
\author{Xin-Jun Ma}
\email{maxiaoguang80@126.com}
\affiliation{College of Mathematics and Physics, Inner Mongolia Minzu University, Inner Mongolia,
Tongliao 028043, China}
\author{Zhi-Qing Li}
\affiliation{Tianjin Key Laboratory of Low Dimensional Materials Physics and Preparing Technology, Department of Physics, School of Science, Tianjin University, Tianjin 300354, China}
\author{Zi-Wu Wang}
\email{ziwuwang@tju.edu.cn}
\affiliation{Tianjin Key Laboratory of Low Dimensional Materials Physics and Preparing Technology, Department of Physics, School of Science, Tianjin University, Tianjin 300354, China}

\begin{abstract}
The self-trapped state (STS) of interlayer exciton (IX) has been aroused enormous interesting owing to their significant impact on the fundamental properties of the van der Waals heterostructures (vdWHs). Nevertheless, the microscopic mechanisms of STS are still controversial. Herein, we study the corrections of the binding energies of the IXs due to the exciton-interface optical phonon coupling in four kinds of vdWHs and find that these IXs are in the STS for the appropriate ratio of the electron and hole effective masses. We show that these STSs could be classified into the type $\mathbb{I}$ with the increasing binding energy in the tens of meV range, which are very agreement with the red-shift of the IXs spectra in experiments, and the type $\mathbb{II}$ with the decreasing binding energy, which provides a possible explanation for the blue-shift and broad linewidth of the IX’s spectra in the low temperature. Moreover, these two types of self-trapped IXs could be transformed into each other by adjusting the structural parameters of vdWHs. These results not only provide an in-depth understanding for the self-trapped mechanism of IX, but also shed light on the modulations of IXs in vdWHs.
\end{abstract}

\maketitle
\section{Introduction}
The interlayer excitons (IXs) are formed by bound pairs of electrons and holes spatially separated in two different two-dimensional (2D) semiconductor materials of van der Waals heterostructures (vdWHs)\cite{d1,d2,d3,d4,d5,d6,d7,d8,d9,d10,d11,d12}, such as those in 2D transition metal dichalcogenides (TMDC) bilayer with a type-II band alignment. The spatial separation between the electron and hole allows achieving long IX lifetimes, orders of magnitude longer than lifetimes of intralayer excitons in TMDC monolayers\cite{d8,d11,d13,d14,d15,d16,d17,d18}. This longer lifetime, combined with the large binding energy, has enabled the exploration of the rich many-body physics of IX and their applications in tunable photonic and optoelectronic devices\cite{d8,d17,d19,d20,d21,d22}.

In the recent years, the self-trapped state (STS) of IX in different vdWHs are arousing more and more attentions because this special state is closely related to the binding energy, lifetime and diffusion length of IX, and thus gives rise to some unique properties of vdWHs. The STS is usually understood to be the result of IX trapped in the potential wells induced by some specific conditions. For example, TMDC bilayer heterostructures with twist-angle induces spatially periodic moir$\acute{e}$ superlattice potential that can trap IX, which has been predicted by several theoretical studies\cite{d23,d24,d25,d26,d27,d28} and confirmed by a series of experiments\cite{d5,d29,d30,d31,d32,d33,d34,d35}. Strain has also been proposed as an effective way to trap IX, in which the band structures of TMDC heterostructures are modified by strain, resulting in the static potential well\cite{d16,d36,d37,d38}. The spatially varying electric fields enable the localization of IX due to its permanent dipole moment\cite{d10,d39,d40}, where the deterministic placement and control of a single trapped IX has been achieved\cite{d39}. Alternatively, free exciton can be trapped by local deformation of the lattice stemming from the strong coupling between excitons and phonons\cite{d52}. In general, these STSs spectra are broaden and have a large Stokes shift\cite{d32,d33,d34,d35,d39}, which may facilitate the improvement of luminescence efficiency of IX. Despite these spectacular discoveries, the microscopic mechanisms for the STS in vdWHs are still under intense debate. In particular, the contribution of the prevalent exciton-phonon coupling on the STS has not yet been well discussed.
\begin{figure*}[htbp]
\centering
\includegraphics[width=7in,keepaspectratio]{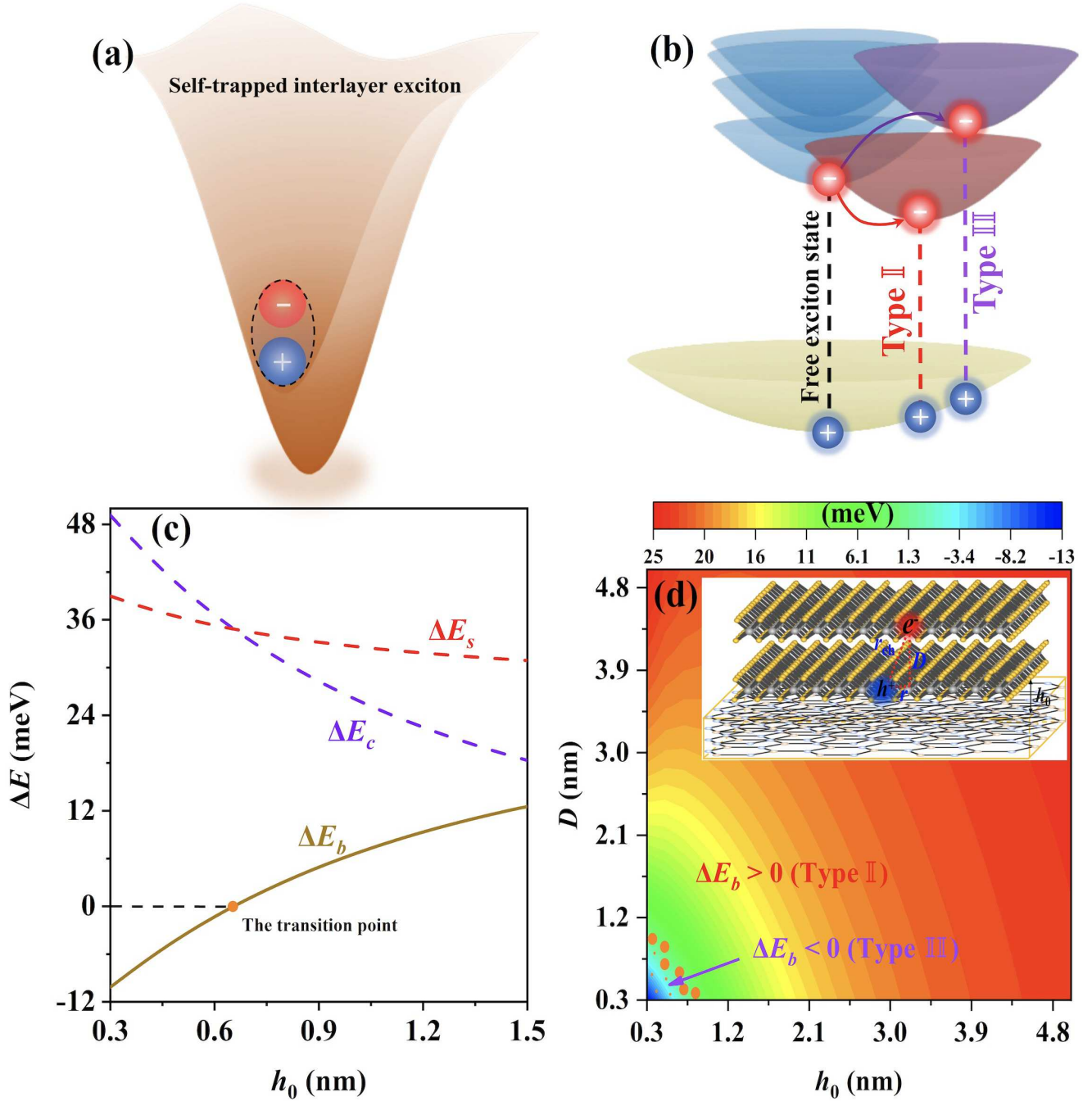}
\caption{\label{compare} (a) The schematic diagram of a self-tapped interlayer exciton (IX) in the potential well generated by lattice distortion. (b) The schematic diagrams of three types of IX: free state, type $\mathbb{I}$ of self-trapped state with the increasing binding energy and type $\mathbb{II}$ of self-trapped state with the decreasing binding energy. (c) The corrections of the self-energy $\Delta$$E_s$, Coulomb potential $\Delta$$E_c$ and binding energy $\Delta$$E_b$ of IX in MoS$_{2}$/MoS$_{2}$/h-BN structure as a function of $h_0$ at $D$ = 0.4 nm. (d) The correction of binding energy ($\Delta$$E_b$) as functions of $h_0$ and $D$.}
\end{figure*}

In the present paper, we study the variations of binding energies of IX ($\Delta$$E_b$), including the self-energy ($\Delta$$E_s$) and Coulomb potential correction ($\Delta$$E_c$), arising from the exciton-interface optical phonon (IOP) coupling for four kinds of vdWHs composed of h-BN and MoS$_2$ (see Figure $S$3 in the Supporting Information), in which different IOP modes and their coupling with IX are given by the dielectric continuum model. We find that these IXs could be in the STS in different heterostructures according to a common criterion $\Delta$$E_s>$ 0, depending on the appropriate ratio of the electron and hole effective mass. Meanwhile, the competition of $\Delta$$E_s$ and $\Delta$$E_c$ can result in the opposite trends of $\Delta$$E_b$, allowing us to divide the STS into two types, namely, type $\mathbb{I}$ with increasing binding energy and the type $\mathbb{II}$ with decreasing binding energy, which indicates that it is insufficient to judge the STS only by the increasing of binding energies in most experiments. Two types of STS could provide the qualitative explanation for some anomalous experimental phenomena in the IX spectra. Furthermore, these two types can be translated into each other by modulating of structural parameters of these heterostructures. These theoretical results enrich the knowledge of the fundamental properties of the self-trapped IX and their modulations in vdWHs.

\section{Results and discussion}
Along with the progresses of the synthesis techniques in 2D heterostructures, controlling the stacking order of 2D materials has become possible in engineering design\cite{d1,d41,d42,d43,d44,d45,d46,d47,d48,d49,d50}, which implies that this bottom-up design can be used to tailor the species of vdWHs. In this work, we mainly study four kinds of vdWHs consisting of h -BN and MoS$_2$: MoS$_2$/MoS$_2$/h-BN, MoS$_2$/h-BN/MoS$_2$/h-BN, h-BN/MoS$_2$/MoS$_2$/h-BN and h-BN/MoS$_2$/h-BN/MoS$_2$/h-BN (see Figure $S$3). The details of the IOP modes and the elements of the exciton-IOP coupling for these heterostructures are given based on the dielectric continuum model (see Section I and II in Supporting Information). The corrections of binding energies ($\Delta$$E_b$) of IX arising from exciton-phonon coupling are calculated using the Lee-Low-Pines unitary transformation method, in which $\widetilde{E_b}$ = $E_{b}$ + $\Delta$$E_{b}$ (see Eq. $S28$) with the intrinsic binding energy of IX ($E_{b}$) (see Eq. $S22$), and $\Delta$$E_b$ is composed of the self-energy ($\Delta$$E_s$) (see Eq. $S31$) and Coulomb potential correction ($\Delta$$E_c$) (see Eq. $S30$) satisfying the relation of $\Delta$$E_b$ = $\Delta$$E_s$ - $\Delta$$E_c$ (see Eq. $S29$). The key parameters for MoS$_2$ and h-BN in numerical calculations are listed in Table $S$1 and evolutions of the IOP modes between MoS$_2$ and h-BN are presented in Figure $S$2 in the Supporting Information. The cut-off wave vectors $q_c$ of these IOP modes could be set to 5 nm$^{-1}$, of which the reliability are discussed in Figures $S$2 and $S$5.

In general, IX generates a lattice distortion potential arising from it coupled strongly with the surrounding phonon bath, and meanwhile, IX is trapped by this potential well\cite{d51,d52}, that is the self-trapping process as shown in Figure 1 (a). In fact, there was a long-lasting problem that how to judge the STS resulting from the exciton-phonon coupling. In previous studies\cite{d53,d56,d57,d58}, $\Delta$$E_s>$ 0, giving the positive contribution to the binding energy, has been widely used as a judging criterion for the STS, where the ratio ($m_e$/$m_h$) of the electron and hole effective mass is required in the range of 0.268 to 3.732. Obviously, for these vdWHs-based MoS$_2$, the ratio $m_e$/$m_h$ = 0.88 ($m_e$ = 0.51 and $m_h$ = 0.58 is commonly employed\cite{d59,d60}) is in this region, so the formed IXs in these vdWHs satisfy this criterion. In the following sections, we will adopt this criterion and present the variations of binding energies of IX for four kinds of vdWHs, respectively. In addition, IXs include A and B types according to the spin selection rules and the spin–orbit interaction between the valence and conduction bands in TMDC materials\cite{d69,d70}. Here, we assume they have the similar behaviors.
\begin{figure}[htbp]
	\includegraphics[width=3.3in,keepaspectratio]{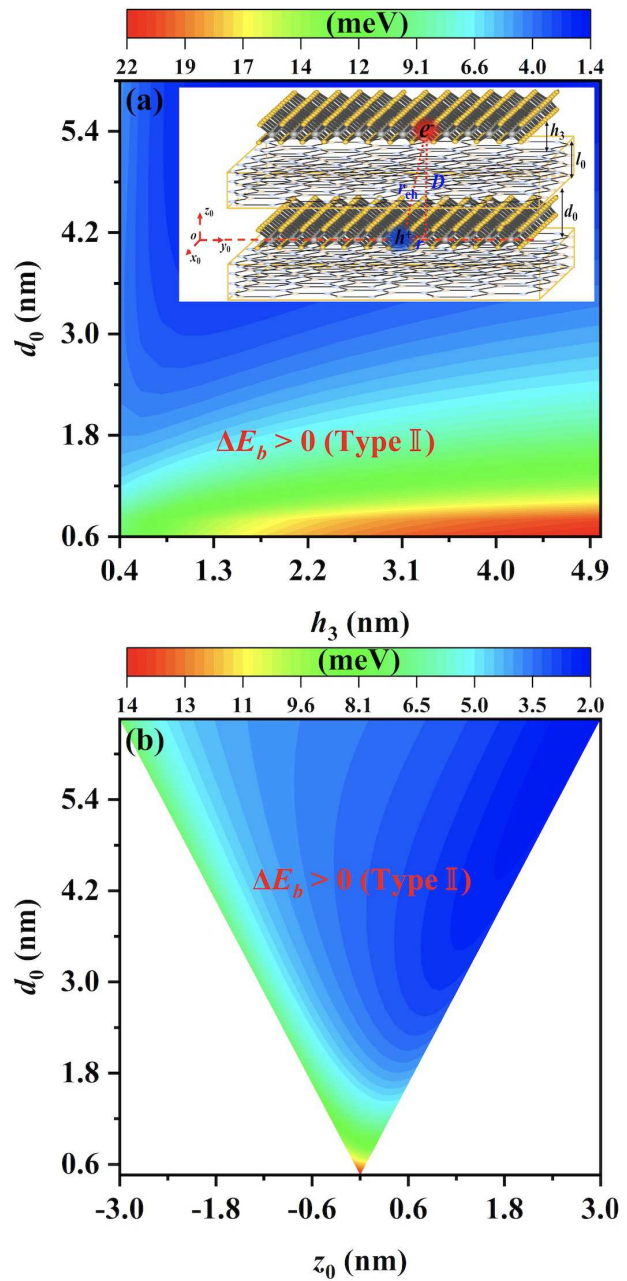}
	\caption{\label{compare} The variation of the binding energy correction $\Delta E_{b}$ of interlayer excitons with respect to internal parameters in MoS$_{2}$/h-BN/MoS$_{2}$/h-BN heterostructure: $l_0$ = 0.333 nm is the thickness of the top h-BN layer. (a) The dependences of $\Delta$$E_b$ on $d_0$ and $h_3$ for the MoS2 layer in the middle of two h-BN encapsulation layers ($z_0$ = 0 nm). (b) The dependences of $\Delta$$E_b$ on $d_0$ and $z_0$ at $h_3$ = 0.6 nm.}
\end{figure}

\subsection{A. MoS$_2$/MoS$_2$/h-BN heterostructure}
As shown the inset in Figure 1 (d) for MoS$_{2}$/MoS$_{2}$/h-BN structure, electron only coupled with the intrinsic longitudinal optical phonons of the top MoS$_{2}$ layer (the coupling element is given by Eq. $S$13), and hole in the bottom MoS$_2$ coupled with the IOP modes between MoS$_2$ and h-BN substrate, depending on the internal distance $h_0$ sensitively (see Eq. $S$14). The Coulomb potential correction ($\Delta$$E_{c}$), self-energy ($\Delta$$E_{s}$) and total correction of binding energy ($\Delta$$E_{b}$) of IX as a function of $h_0$ are shown in Fig. 1 (c). One can see that $\Delta$$E_{c}$ decays more faster than $\Delta$$E_{s}$ with increasing $h_0$, which results in $\Delta$$E_{b}$ undergoing a transition from negative to positive ($\Delta$$E_{b}$ = $\Delta$$E_{s}$ - $\Delta$$E_{c}$). 
This allows the STS to be divided into two types: type $\mathbb{I}$ with the energy shifting downwards from the free exciton state (FE) and type $\mathbb{II}$ with the energy shifting upwards, as illustrated in Figure 1 (b), indicating that it is insufficient to judge the STS only by the increasing of binding energy in most experiments\cite{d34,d80,d81,d82,d83}. Except the internal distance $h_0$, the spatial distance $D$ between electron and hole has a direct impact on binding energy since it determines the Coulomb interaction of electron-hole pair and thus the variation of $\Delta$$E_{b}$, as shown in Figure 1 (d). We can see that $\Delta$$E_b$ varies from -13 meV to 25 meV, in which the regions of types $\mathbb{I}$ and $\mathbb{II}$ are clearly marked. A series of recently experiments have shown that the IX photoluminescence peak is red-shifted by tens of meV relative to its absorption peak in the experiments\cite{d36,d43,d44,d45,d46,d61,d62}, which is very consistent with the increasing magnitude of the binding energy of type $\mathbb{I}$ (0 $\sim$ 25 meV). This type of STS could be thermally excited to the free exciton, which may give an explanation that the narrowing linewidth of the IX photoluminescence spectrum with increasing temperature in TMDC heterostructures\cite{d68}. Compared to type $\mathbb{I}$, the observation of type $\mathbb{II}$ will be extremely challenging in the experiments owing to facts that (1) it is confused with the excited states of free exciton as schemed in Figure 1 (b); (2) the region of type $\mathbb{II}$ is much smaller than that of type $\mathbb{I}$. Moreover, the type $\mathbb{II}$ will disappear with increasing $h_0$ and $D$, because the Coulomb interaction of electron-hole pair and the strength of hole-IOP coupling decrease obviously with $D$ and $h_0$, giving rise to the decay of $\Delta$$E_c$ being much faster than $\Delta$$E_s$, which further enhances the difficulty of the distinguish for the type $\mathbb{II}$ of STS. Nevertheless, at low temperature, the blueshift and broadened linewidth of the IX peak have been observed experimentally in TMDC heterostructures\cite{d68}, which was explained by Karni $et$ $al.$ as possibly arising from some local states, such as the moir$\acute{e}$ pattern, defects, and impurities. It is supposed from our results that type $\mathbb{II}$ of STS with the decreasing binding energy may also provide a potential explanation for these abnormal phenomena. In fact, these peculiar properties of IX are also appeared in the following structure of h-BN/MoS$_2$/MoS$_2$/h-BN in Figure 3.

\subsection{B. MoS$_2$/h-BN/MoS$_2$/h-BN heterostructure}
The bottom MoS$_2$ layer is encapsulated to form MoS$_2$/h-BN/MoS$_2$/h-BN heterostructure (the inset in Figure 2 (a)), in which the coupling element (see Eq. $S$15) between the electron and the IOP modes is sensitive to $h_3$ (the distance between the top h-BN layer and the top MoS$_2$ layer) and the coupling element (see Eq. $S$16) between the hole and the IOP modes depends on $z_0$ (representing the position of MoS$_2$ between h-BN bilayer) and $d_0$ (the distance between h-BN bilayer). The dependences of $\Delta$$E_{b}$ on these structural parameters are shown in Figure 2. In Figure 2 (a), $\Delta$$E_b$ varies from 1.4 meV to 22 meV with $h_3$ and $d_0$, which means that IXs are only in the type $\mathbb{I}$ states in this structure. 
\begin{figure}[t]
	\includegraphics[width=3.4in,keepaspectratio]{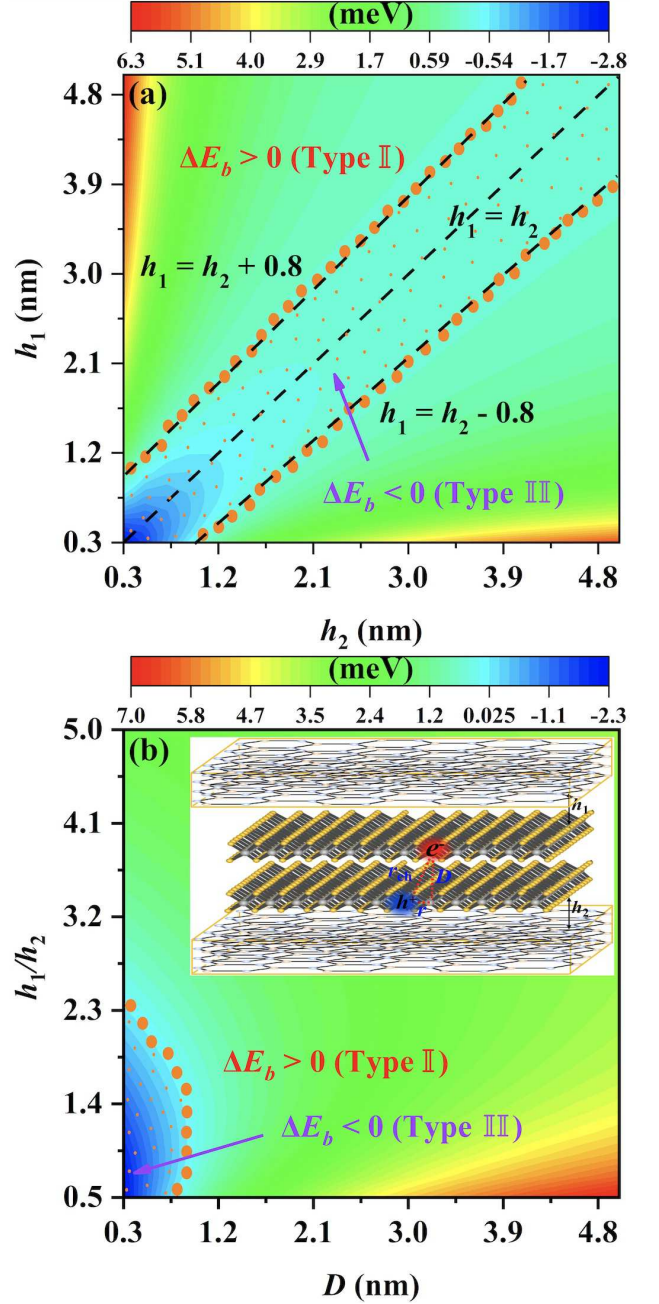}
	\caption{\label{compare} The corrections of the binding energy $\Delta E_{b}$ of interlayer excitons for h-BN/MoS$_{2}$/MoS$_{2}$/h-BN heterostructure. (a) $\Delta E_{b}$ as functions of $h_1$ and $h_2$ at $D$ = 0.4 nm. (b) $\Delta E_{b}$ as functions of the ratio of $h_1$/$h_2$ and $D$ ($h_2$ = 0.6 nm is assumed).}
\end{figure}
This because both the IX-IOP coupling and the Coulomb interaction of electron-hole pair are severely weakened by increasing $h_3$ and $d_0$, rendering the variation of $\Delta$$E_{c}$ being always less than $\Delta$$E_{s}$. Anther remarkable feature is that the correction effect of $d_0$ on the binding energy is more stronger than $h_3$, which can be attributed to the encapsulation layers providing three effective IOP modes that coupled with the hole (see Figure $S$2). In practice, the position ($z_0$) of MoS$_2$ layer may fluctuate between two h-BN encapsulation layers depending on the fabrication methods in engineering designs for heterostructures, which inevitably imposes the influence on the IX binding energy, as shown in Figure 2 (b). We find that $\Delta$$E_b$ varies asymmetrically from 2.0 to 14 meV with respect to $z_0$. In fact, the strength of hole-IOP coupling is symmetrically dependent on the variation of $z_0$ as given by Eq. $S$16 (the elelctron-IOP coupling does not depend on $z_0$), and thus the evolution of $\Delta$$E_s$ (see Eq. $S$31) is symmetrical. But the spacing distance $D$ between the electron and hole varies with $z_0$, which modifies the strength of the Coulomb interaction between them substantially, giving rise to the asymmetrical variation of $\Delta$$E_c$ (see Eq. $S$30). Consequently, the asymmetry of $\Delta$$E_b$ ($\Delta$$E_b$ = $\Delta$$E_s$ - $\Delta$$E_c$) is formed as plotted in Figure 2 (b). Unfortunately, these theoretical results for this structure cannot be compared with experiments until now because of lack of corresponding data.
\begin{figure}
	\centering
	\includegraphics[width=3.3in,keepaspectratio]{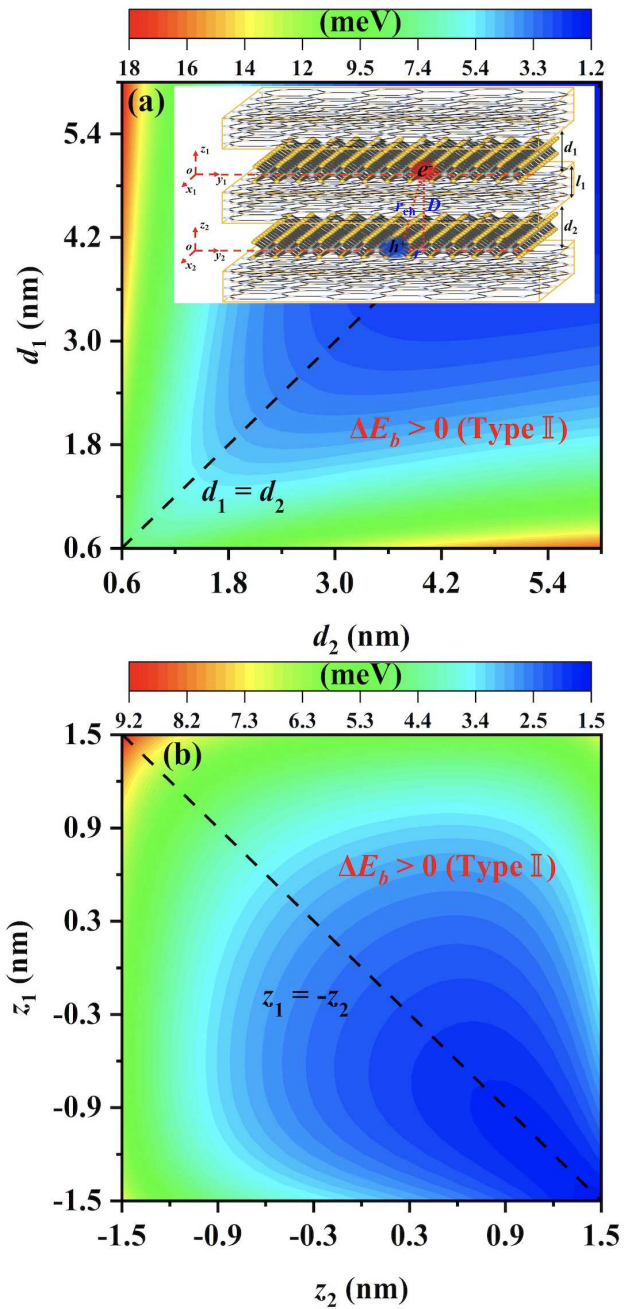}
	\caption{\label{compare} The binding energy correction $\Delta E_{b}$ of interlayer excitons with respect to internal parameters in h-BN/MoS$_2$/h-BN/MoS$_2$/h-BN heterostructure. (a) The relations of $\Delta E_{b}$ with $d_1$ and $d_2$ for the MoS$_2$ layer being in the middle of two h-BN layers (that is $z_1$ = $z_2$ = 0 nm being assumed). (b) The relations of $\Delta E_{b}$ with $z_1$ and $z_2$ at $d_1$ = $d_2$ = 3.46 nm. $l_1$ = 0.333 nm for the thickness of the intermediate h-BN layer.}
\end{figure}

\subsection{C. h-BN/MoS$_2$/MoS$_2$/h-BN heterostructure}
The h-BN/MoS$_2$/MoS$_2$/h-BN (the inset in Figure 3 (b)) was the extensive used heterostructure in a large number of experiments\cite{d32,d33,d34,d39,d83,d73}, in which the atomically thin h-BN is an ideal gate material with the high resistance to both mechanical manipulation and chemical reactions to ensure the stability of heterostructures\cite{d47,d75,d76,d77}.
 In this configuration, the strength of the electron (hole)-IOP coupling sensitively depends on $h_1$ ($h_2$), described by Eq. $S$17 ($S$18), where $h_1$ ($h_2$) is the distance between the top (bottom) h-BN layer and the top (bottom) MoS$_2$ layer. In Figure 3 (a), it can be see that $\Delta$$E_b$ varies nearly symmetrically with respect to $h_1$ = $h_2$. In the symmetrical region bounded by $h_1$ = $h_2$ + 0.8 nm and $h_1$ = $h_2$ - 0.8 nm, the type $\mathbb{II}$ of the STS is emerged originating from the strength of the electron-IOP coupling being nearly equivalent to that of the hole, thus resulting in $\Delta$$E_c$ prevailing over $\Delta$$E_s$. On the contrary, the type $\mathbb{I}$ of the STS is formed outside this region, where the difference of the coupling strengths becomes obvious. We also present the influence of the distance ($D$) between two MoS$_2$ layers on $\Delta$$E_b$ for different ration of $h_1$/$h_2$ in Figure 3 (b). One can see that the STS transits from the type $\mathbb{II}$ to $\mathbb{I}$ with increasing $D$, which attributes to the changing magnitude of $\Delta$$E_c$ compares to that of $\Delta$$E_s$ as $D$ increases. This sheds light on a potential way to control the phase transitions between two types of STS by manipulating $h_1$, $h_2$ and $D$. Furthermore, comparing with the heterostructures in Figures 1 and 2, the variational magnitude of the binding energy of STS is very small in this structure (less than ten meV), implying that three types of IXs (free exciton, types $\mathbb{I}$ and $\mathbb{II}$) are easily transformed into each other, which possibly causes that, on the one hand, the distinguish between them is extremely difficult; on the other hand, the luminescence efficiency of the IX is significantly improved, which could be used to explain the experimental observations that the IX emission lines (lifetime) in the encapsulated heterostructures are more wider\cite{d68} (longer\cite{d78,d79}) than the unencapsulated ones.    

\subsection{D. h-BN/MoS$_2$/h-BN/MoS$_2$/h-BN heterostructure}
The fourth type of heterostructure (the inset in Figure 4 (a)) is a fully encapsulated structure, in which $d_1$ ($d_2$) is the distance between the top (bottom) h-BN bilayer and $z_1$ ($z_2$) serves to describe the change in position of top (bottom) MoS$_2$ between h-BN bilayer. The coupling element between electron and IOP modes, depending on $d_1$ and $z_1$, is given by Eq. $S$19 and the hole-IOP coupling, depending on $d_2$ and $z_2$, is described by Eq. $S$20. The evolutions of $\Delta$$E_b$ with respect to the structural parameters $d_1$ and $d_2$ as well as $z_1$ and $z_2$ are shown in Figures 4 (a) and (b), respectively. One can see that $\Delta$$E_b$ is symmetrical respect to $d_1$ = $d_2$ and $z_1$ = -$z_2$, and only the type $\mathbb{I}$ of STS is emerged, which similar to MoS$_2$/h-BN/MoS$_2$/h-BN heterostructure in Figure 2. Therefore, it could be concluded that the full encapsulation of MoS$_2$ layer suppresses the emergence of the type $\mathbb{II}$ of STS in the vdWHs. In Figure 4 (a), $\Delta$$E_b$ varies in the range of 1.2 - 18 meV, where the minimal values of the type $\mathbb{I}$ of STS are at $d_1$ = $d_2$. This also can be attributed to the difference between the strengths of the electron- and hole-IOP coupling in this condition is the smallest. Furthermore, Figure 4 (b) shows an asymmetrical evolution of $\Delta$$E_b$ with respect to $z_1$ and $z_2$. According to Eqs. $S$19 and $S$20, the elements of electron- and hole-IOP coupling vary symmetrically with $z_1$ and $z_2$, respectively, which will result in $\Delta$$E_s$ having the same changing trend. Meanwhile, the variations of $z_1$ and $z_2$ also affect the spatial distance $D$ between the electron and hole directly, which gives rise to $\Delta$$E_c$ changing asymmetrically. As a result, the competition between $\Delta$$E_s$ and $\Delta$$E_c$ ($\Delta$$E_b$ = $\Delta$$E_s$ - $\Delta$$E_c$) leads to the asymmetrical trend of $\Delta$$E_b$. Nevertheless, the validity of these results is limited by lack of the relevant experimental reports. We also expect these theoretical predictions  stimulate more experiments in this aspect.

\section{CONCLUSION}
In summary, we have revealed the variations of the binding energies of the IXs owing to exciton-phonon coupling in four kinds of vdWHs. We found that the STS could be divided into two types: type $\mathbb{I}$ with increasing binding energy and type $\mathbb{II}$ with decreasing binding energy, in which the corrected amplitudes of the binding energies and the phase transitions between them are determined by the Coulomb interaction of the IX and the difference between the strengths of the electron- and hole-IOP coupling, both of which can be effectively tuned by the structural parameters of these heterostructures. Furthermore, the varying stacking sequence of vdWHs is beneficial for maintaining the stability of the IX, such as the h-BN/MoS$_2$/MoS$_2$/h-BN heterostructure is the most stable heterostructure, and the full encapsulation of single or double MoS$_2$ layers could suppress the emergence of type $\mathbb{II}$ of STS. These point us to some potential pathways for tailoring the expected IXs in the engineering design of vdWHs.

 \section{Acknowledgement}
This work was supported by National Natural Science Foundation of China (Grand Nos. 12174283 and 12164032) and Scientific Research Project of Institutions of Higher Learning in Inner Mongolia Autonomous Region (Grant No. NJZY22137).

\section{Data Availability Statement}
The data that supports the findings of this study are available within the article.

\end{document}